\documentclass[conference]{IEEEtran}

\usepackage{imakeidx}
\makeindex
\usepackage{cite}
\usepackage{placeins}
\usepackage{url}
\usepackage[dvipsnames]{xcolor}
\usepackage{dutchcal}
\usepackage{graphicx}
\usepackage{subcaption}

\usepackage{algorithm}
\usepackage{algpseudocode}
\usepackage{amsmath}
\usepackage{amssymb}

\usepackage{tikz}
\usetikzlibrary{shapes.geometric, arrows.meta, positioning}

\tikzstyle{process} = [rectangle, minimum width=3.5cm, minimum height=1.2cm, text centered, draw=black,align=center, fill=blue!10]
\tikzstyle{datablock} = [trapezium, trapezium left angle=70, trapezium right angle=110, minimum height=1cm, text centered, draw=black, align=center, fill=orange!20]
\tikzstyle{arrow} = [thick,->,>=Stealth]

\usepackage{array}      
\usepackage{booktabs} 
\graphicspath{ {./images/} }
\ifCLASSINFOpdf

\else

\fi

\begin{document}

\title{Toward a Universal Color Naming System: A Clustering-Based Approach using Multisource Data}
\author{\IEEEauthorblockN{Aruzhan Sabitkyzy, Maksat Shagyrov, Pakizar Shamoi*}
\IEEEauthorblockA{School of Information Technology and Engineering\\ Kazakh-British Technical University\\
Almaty, Kazakhstan\\
*Email: p.shamoi@kbtu.kz 
}
}

\maketitle

\IEEEpeerreviewmaketitle

\begin{abstract}


Is it coral, salmon, or peach? What seems like a simple color can have many names, and without a standard, these variations create confusion across design, technology, and communication.
Color naming is a fundamental task across industries such as fashion, cosmetics, web design, and visualization tools. However, the lack of universally accepted color naming standards leads to inconsistent color standards across platforms, applications, and industries. Moreover, these systems include hundreds or thousands of overlapping, perceptually indistinct shades, despite the fact that humans typically distinguish only a limited number of unique color categories in practice. In this study, we propose a clustering-based multisource data framework to build a standardized color-naming system. We collected a dataset of over 19,555 RGB values paired with color names from 20 diverse sources. After data cleaning and normalization, we converted the colors to the perceptually uniform CIELAB color space and applied K-means clustering using the CIEDE2000 color difference metric, identifying 280 optimal clusters. For each cluster, we performed a frequency analysis of the associated names to assign representative labels. The resulting system reflects naturally occurring linguistic patterns. We demonstrate its effectiveness in automatic annotation and content-based image retrieval on a clothing dataset. This approach opens new opportunities for standardized, perceptually grounded color labeling in practical applications such as generative AI, visual search, and design systems.


\end{abstract}
\section{Introduction}
Color is a fundamental aspect of human perception that shapes how we experience and communicate with our environment. It influences a wide range of disciplines, including psychology, physics, engineering, and computer science \cite{elliot2014color, nassau2001physics, plataniotis2000color}. Color information is essential for image processing\cite{plataniotis2000color}, human-computer interaction\cite{yang1998visual}, and its features facilitate object recognition by enabling segmentation based on visual properties\cite{gevers2000pictoseek}, thereby making object detection\cite{gevers1999color}, scene interpretation, and decision-making automation easier.

There are two main ways to specify color: numerical codes, e.g., RGB and HEX representations, and linguistic descriptors, that is, color names. While numerical codes precisely specify color composition, they are unsuitable for everyday communication. Color naming, by contrast, is the process of assigning intuitive, easy-to-read names to colors according to their appearance under some perception conditions \cite{menegaz2006discrete}. Color naming is formally defined as a function “N” that translates visual stimuli into color terms \cite{lammens1994computational}.

 Color naming is important in real-world applications to connect the human user and the machine system. Companies use color names to gain consumers' attention \cite{cr5} and provide recommendations of aesthetic color palettes \cite{10001872}. On e-commerce websites, color search \cite {smith1997visualseek} is a valuable functionality for product exploration, with large sites such as Amazon using color filtering to help customers find products that match their desired colors. In user interface design, it would also be impractical to use color codes, such as RGB values, because end users prefer descriptive color names over technical jargon.

 Despite the significance of color naming, there is a critical problem: the lack of a universally accepted standard for color naming.  Color naming is sensitive and subjective. Languages and cultures differ in how they categorize and name colors.  Various companies, industries, and platforms have their own naming systems, which leads to misunderstandings and miscommunication, leading to varied interpretations of color names, e.g., olive, terracota, etc.

\begin{figure}[tb]
\centering
\begin{minipage}[t]{0.49\linewidth}
    \centering
    \includegraphics[width=\linewidth]{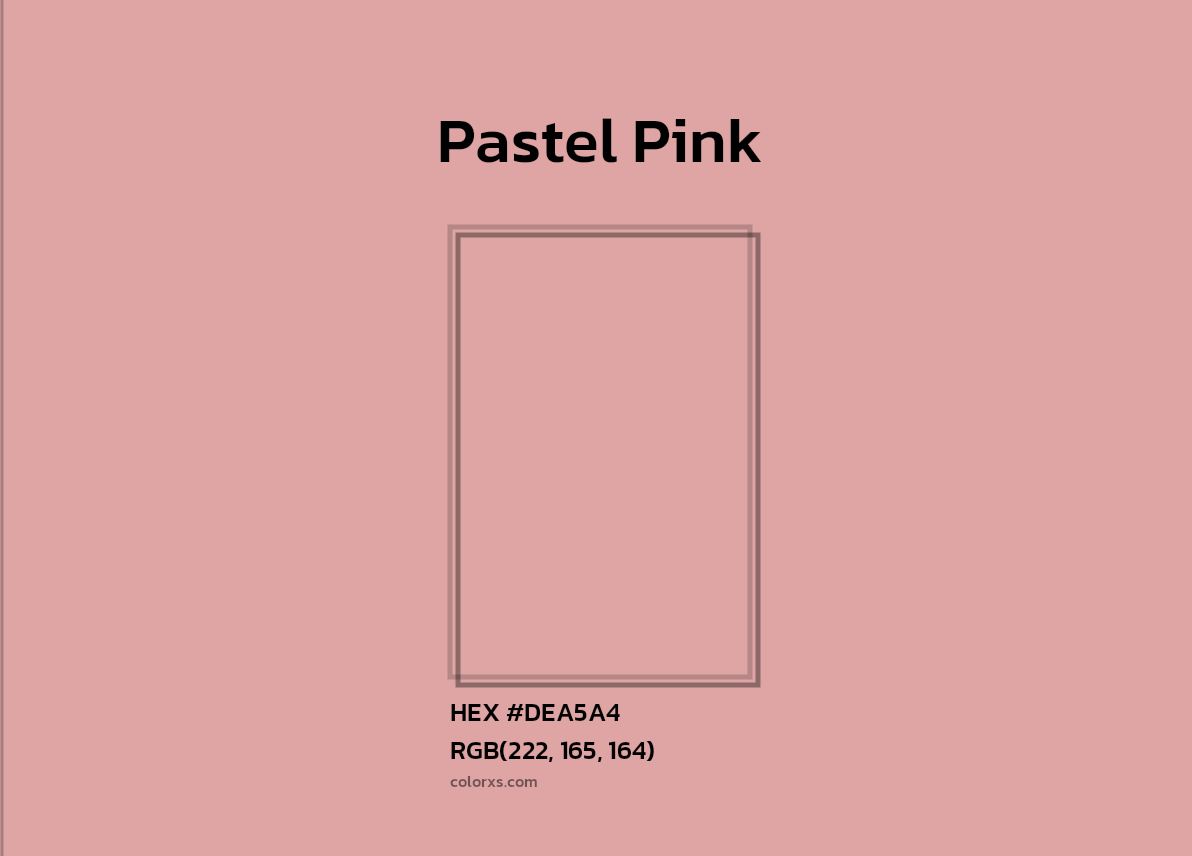}
    \subcaption{colorsx.com} \label{fig:top-left}
\end{minipage}
\hfill
\begin{minipage}[t]{0.49\linewidth}
    \centering
    \includegraphics[width=\linewidth]{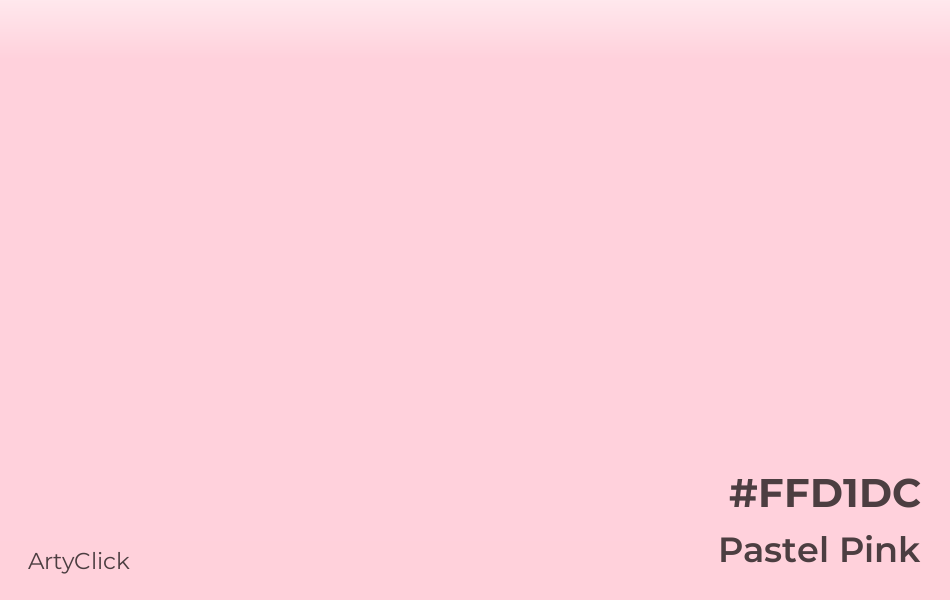}
   \subcaption{artyclick.com} \label{fig:top-right}
\end{minipage}

\vspace{1em}

\begin{minipage}[t]{0.49\linewidth}
    \centering
    \includegraphics[width=\linewidth]{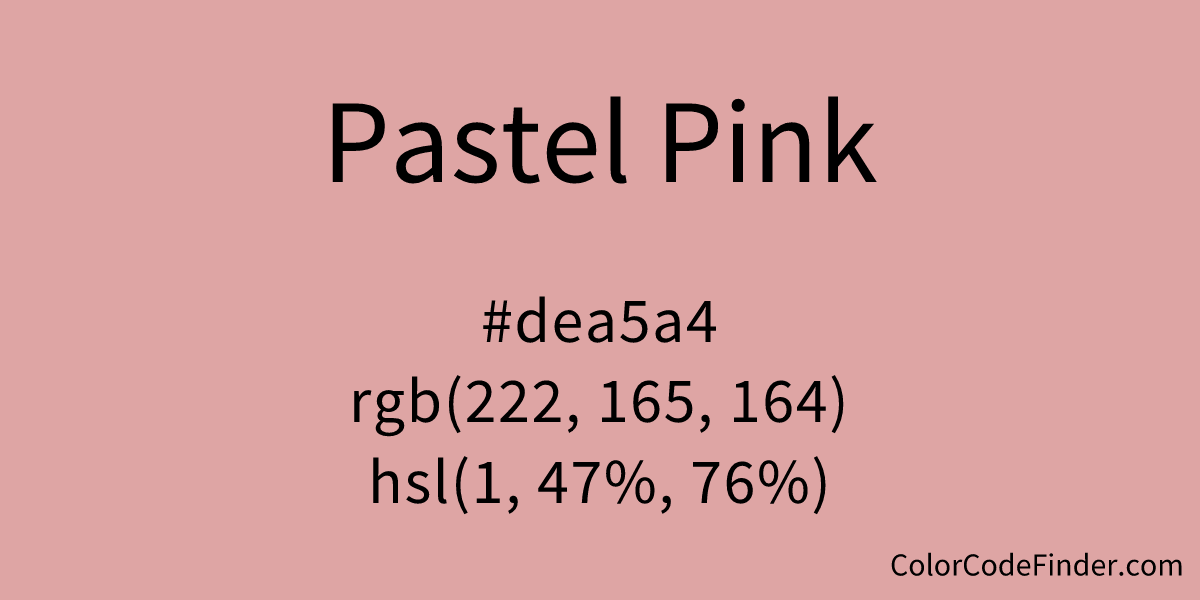}
    \subcaption{colorcodefinder.com} \label{fig:bottom-left}
\end{minipage}
\hfill
\begin{minipage}[t]{0.49\linewidth}
    \centering
    \includegraphics[width=\linewidth]{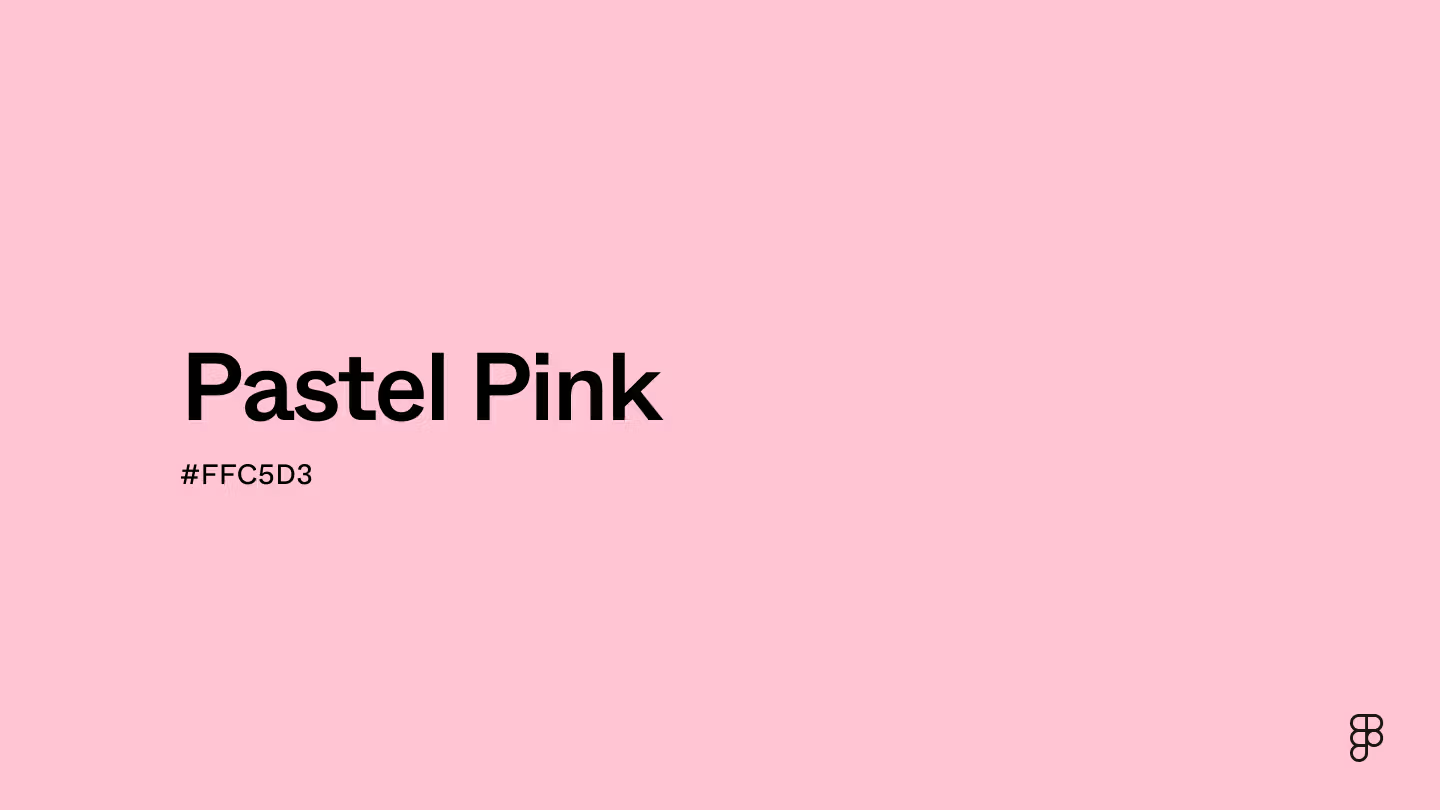}
    \subcaption{figma.com} \label{fig:bottom-right}
\end{minipage}

\caption{Inconsistency of color name across platforms}
\label{fig:image-grid}
\end{figure}

 For example, one company's 'pastel pink' can be quite different from another's - even if the names are identical - since their RGB and hex codes can also be different, as can be seen in Fig. \ref{fig:image-grid}. It presents four varying 'pastel pinks'  - \ref{fig:top-left}: from colorxs.com; \ref{fig:top-right}: from artyclick.com; \ref{fig:bottom-left}: from colorcodefinder.com; and \ref{fig:bottom-right}: from figma.com. Even sophisticated search engines struggle to resolve these discrepancies when matching user requests to visual content.

While there are variarions in the number and boundaries of color terms, there are strong universal tendencies in how colors are categorized and named. Most languages cluster color names into a limited set of categories (e.g., black/white, then red, then green/yellow, etc.) \cite{Kay2003Resolving, Loreto2012On, Lindsey2006Universality, Baronchelli2009Modeling }.

 The lack of standardization poses significant challenges for information retrieval, e-commerce, interface design, and cross-disciplinary communication. Industries rely on distinct color interpretations, often leading to inconsistencies. It makes the demand for more systematic approaches to color naming increasingly urgent. Not only is such a system essential for enhancing clarity and consistency across platforms, but it will also be required to enable future technologies. Generative AI\cite{feuerriegel2024generative} models, for example, such as those employed in text-to-image generation or design automation, depend heavily on well-mapped semantic color data\cite{li2019controllable}. Without standardized names, these models may provide inconsistent or incorrect images. Color naming standardization can also enhance accessibility tools for users with color vision deficiency, improve dataset annotation for machine learning deployment, and facilitate cross-linguistic communication for globally deployed applications.

We propose a standardized color-naming system based on clustering 19555 color samples (RGB values and names) collected from 20 diverse sources. After data cleaning and text normalization, we applied K-means clustering (optimal k=280, determined using the Elbow method) in the perceptual LAB color space, using the CIEDE2000 distance metric. For each cluster, we analyzed the most frequent color names to assign representative labels. The resulting system supports consistent color naming and demonstrates practical benefits in automatic image annotation tasks, such as content-based retrieval and labeling in a clothing dataset. To the best of our knowledge, this is the first data-driven attempt to develop a comprehensive color-naming system.
 
The key contributions of this paper are: 
\begin{itemize}
\item Building a large dataset with more than 19,555 RGB values with names collected from 20 sources of colors; 
\item Defining a standardized color naming system consisting of 280 colors by assigning to each RGB value multiple color names?
\item Application example in CBIR

\end{itemize}

The paper is structured as follows. This section is the Introduction. Section II provides an overview of related works. Next, Section III describes methods, including data collection, k-means clustering and proposed approach. Results are presented in Section IV. Finally, Section V provides concluding remarks of the study.

\section{Related Work}
\subsection{Standardized Frameworks and Systems}


Standardized color frameworks have long aimed to unify how colors are named, classified, and communicated across scientific, industrial, and design contexts. Nowadays, a number of color naming frameworks exist.

The Universal Color Language (UCL) and ISCC–NBS Method provide hierarchical naming schemes that bridge perceptual categories with scientifically defined color boundaries, enabling consistent use of color terminology across disciplines \cite{Kelly2018Color}.

Next, the Natural Color System (NCS) offers a perceptually based notation grounded in Hering’s opponent-color theory. It treats color as a visual experience and provides a structured symbolic language, supported by the NCS Color Atlas, widely used in design, architecture, and industrial color specification\cite{Hard1981NCS—Natural}.

The Munsell Color System organizes color through hue, value, and chroma in a perceptually uniform structure. Although originally based on controlled color chips, a later study shows that color names learned from real-world images can outperform chip-based naming in applied tasks, revealing differences in chromatic and achromatic classifications when Munsell arrays are interpreted using data-driven models \cite{VanDeWeijer2009Learning}.

Furthermore, the RAL Design System defines over 2,500 standardized colors, each with a numeric code and descriptive name (e.g., RAL 9010 Pure White), and is widely used in architecture, manufacturing, and product design. Research demonstrates how Munsell colors can be mapped to RAL equivalents, highlighting the practical need for cross-system conversion and consistent nomenclature \cite{ozturk2005}. Together, these systems, UCL/ISCC–NBS, NCS, Munsell, and RAL Design, represent complementary approaches to color standardization, combining hierarchical naming and perceptual modeling to support clear and consistent color communication across diverse applications.
 
Beyond open and perceptually grounded frameworks, color standardization is also addressed through industry-specific and digital conventions. Pantone \footnote{https://www.pantone.com/} is widely adopted in graphic design, printing, and fashion, providing standardized named colors to ensure consistent color reproduction across physical media. By contrast, HTML and web-based color specifications define colors using numeric RGB and hexadecimal representations, enabling precise digital rendering across devices while offering limited semantic naming and minimal perceptual grounding.

\subsection{Computational Color Naming System}


 Color naming has developed significantly as a research discipline in the last few decades, from simple linguistic investigations to complex computational models. Mapping linguistic color names onto color values has been automated and formalized. 
 

 The systematic study of color naming dates back to Berlin and Kay's foundational work \cite{berlin1991basic}, in which 11 universal color terms were established: white, black, red, green, yellow, blue, brown, pink, purple, orange, and gray. This foundational work established a model that would serve as a conceptual foundation for future computational methods. Another study \cite{boynton1987locating} introduced foundational experiments that mapped these basic color terms to specific regions of color space, creating the first datasets usable for computational modeling.

The development of computational color naming systems primarily focused on establishing direct mappings between color coordinates and linguistic labels. One of the first computational methods was implemented as dictionaries of color names, mapping specific RGB values to basic color terms \cite{berk1982human}. These early systems were based on rigid boundaries in color space, assigning each point to precisely one color category. Building on this work, a more complex computational framework was proposed that incorporates color naming and composition color descriptors, showing the relationship with images \cite{mojsilovic2005computational}.

 Later on, researchers began developing probabilistic models capable of quantifying the inherent uncertainty in human color naming. Mylonas\cite{mylonas2010towards} employed probabilistic algorithms for mapping color values onto color names in cooperation with human observers. Another group of researchers\cite{heer2012color} suggested utilizing multinomial probability distributions for modeling how the color naming model could enhance name-based pixel picking in such applications as image editing, color dictionaries \& thesaurus, and evaluation tools for comparing color palette designs.

The fuzzy approach has received wide approval \cite{kay1978linguistic}. Additionally, comparative experiments between fuzzy logic models and prototype theory models of computational color naming revealed that fuzzy models are more adaptable and more accurately represent human color perception than prototype theory models \cite{moroney2008lexical}. As noted in the paper \cite{benavente2008parametric}, fuzzy models can be improved by proposing a set of parameters that minimize errors in fitting both the training and test datasets. Lastly, a fuzzy model was introduced that mimics how individuals perceive color differences and similarities provides a more comprehensive understanding of color perception \cite{seaborn2005fuzzy}.

As we can see, there are several limitations of existing color standards, such as an excessive number of colors, overlapping shades with different names, and difficulty in human perception.

Although numerous computational models of color naming have been developed, none have directly addressed the practical challenge of consolidating various naming systems into a standardized system. This work addresses this gap by proposing a standardized computational color naming system based on fuzzy string matching and selection from a set of canonical names. The system is based on human perception limits, which would serve as a reliable, cross-industry standard for color names.


Studies have suggested that our visual system can distinguish about 2 to 10 million colors \cite {judd1975color, hardin1992virtues, goldstein1996sensation}. Although we are able to perceive more than two million different colors, colors are commonly grouped into a number of more or less discrete categories
The process of color categorization demonstrates how perceptual boundaries become linguistic boundaries. As noted in cross-cultural studies, when people are required to assign names to colors, perceptually similar colors fall into the same categorical groupings regardless of their distinct RGB values. For example, multiple shades that might be technically different can all be categorized under a generalized name like "violet" because they appear sufficiently similar to human observers. This phenomenon reflects what Berlin and Kay \cite{berlin1991basic} identified as the universal tendency to partition continuous color space into discrete lexical categories.

Widely used color order systems, such as the Munsell Color System and the Natural Color System (NCS), were primarily developed using expert-driven and geometric principles rather than large-scale empirical data derived from real-world color usage.\cite{Hard1981NCS—Natural}\cite{VanDeWeijer2009Learning} While these systems provide perceptually coherent representations of color space, they largely focus on perceptual geometry and make little use of bottom-up aggregation from natural language data or large-scale crowdsourced color naming.

\section{Methodology}
\subsection{Proposed Approach}

The proposed approach combines computational color analysis and natural language processing techniques. 

\begin{figure}[h!]
\centering
\begin{tikzpicture}[node distance=0.5cm and 0.5cm]

\node (data) [datablock] {19,555 Color Samples \\ (RGB + Names) from 20 Sources};
\node (clean) [process, below=of data] {Data Cleaning \& Text Preprocessing};
\node (clustering) [process, below=of clean] {Clustering in LAB Space \\ (K-means, $k=280$, CIEDE2000)};
\node (naming) [process, below=of clustering] {Frequency Analysis \\ of Color Names per Cluster};
\node (output) [datablock, below=of naming] {Standardized Color Naming \\ System (280 Colors)};
\node (application) [process, below=of output] {Applications: \\ CBIR, Auto-Annotation, GANs};

\draw [arrow] (data) -- (clean);
\draw [arrow] (clean) -- (clustering);
\draw [arrow] (clustering) -- (naming);
\draw [arrow] (naming) -- (output);
\draw [arrow] (output) -- (application);

\end{tikzpicture}
\caption{Overview of the proposed data-driven color naming system.}
\label{fig:color_pipeline}
\end{figure}
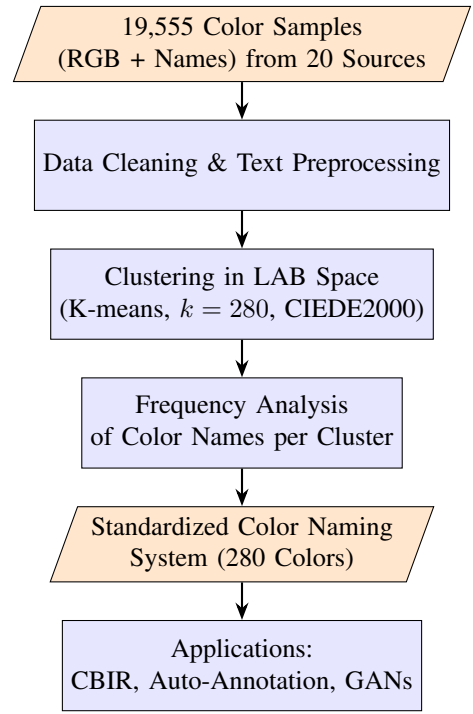

\begin{figure*}[tb]
    \centering
    \includegraphics[width=0.9\textwidth]{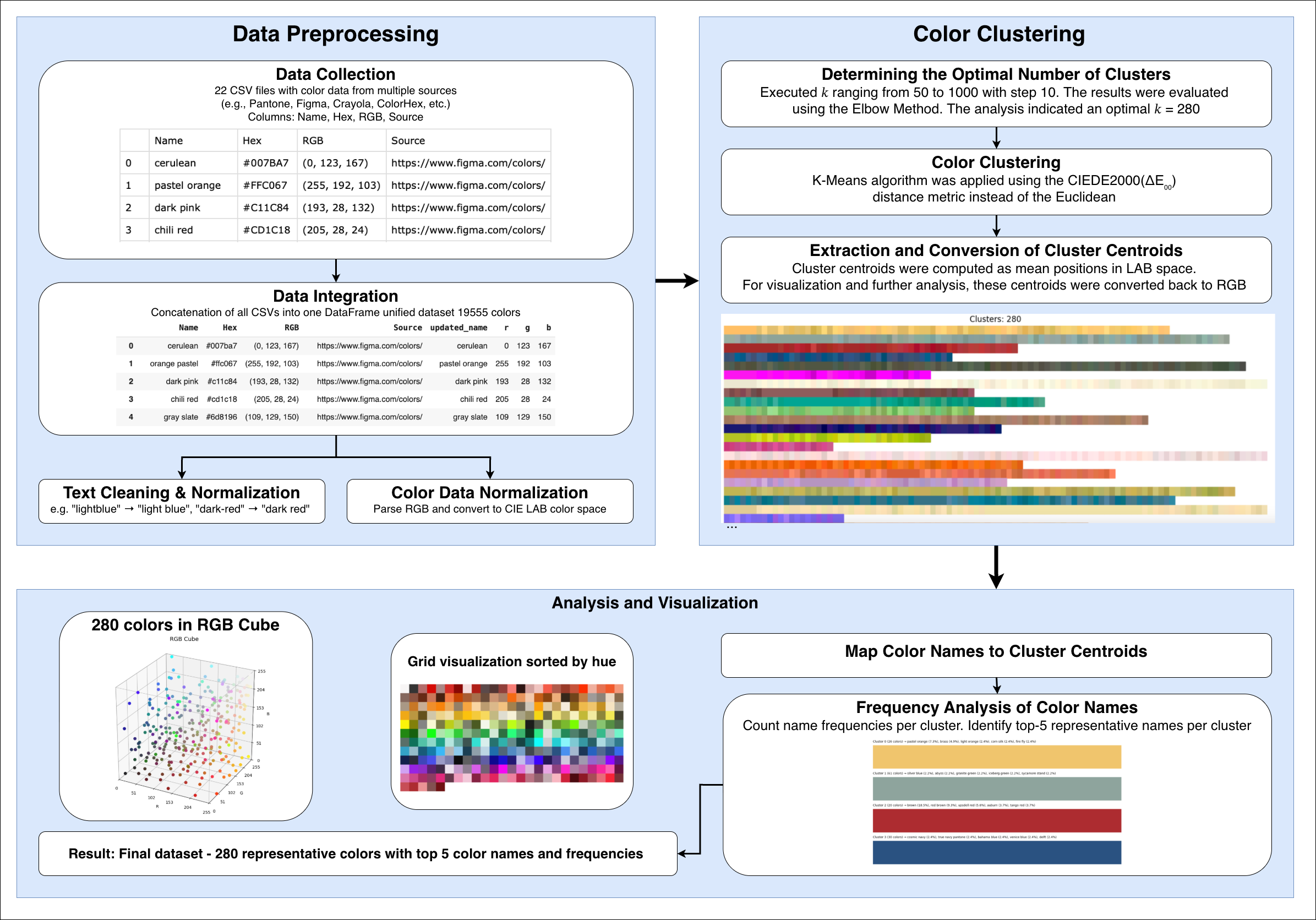}
     \caption{Methodology of the study}
    \label{fig:main}
\end{figure*}

The pipeline of the suggested data-driven color naming system is shown in Fig. \ref{fig:color_pipeline}, which begins with 19,555 RGB color samples with corresponding names gathered from 20 sources. Next, the data is cleaned and clustered in CIELAB space using k-means with the CIEDE2000 distance metric. Frequency analysis is used to assign standardized color names to the generated clusters, yielding a 280-color naming system suitable for applications such as CBIR, automatic annotation, and GANs. The detailed methodology is presented in Fig. \ref{fig:main}.



 \subsection{Data}
 \begin{figure*}
    \centering
    \includegraphics[width=0.7\linewidth]{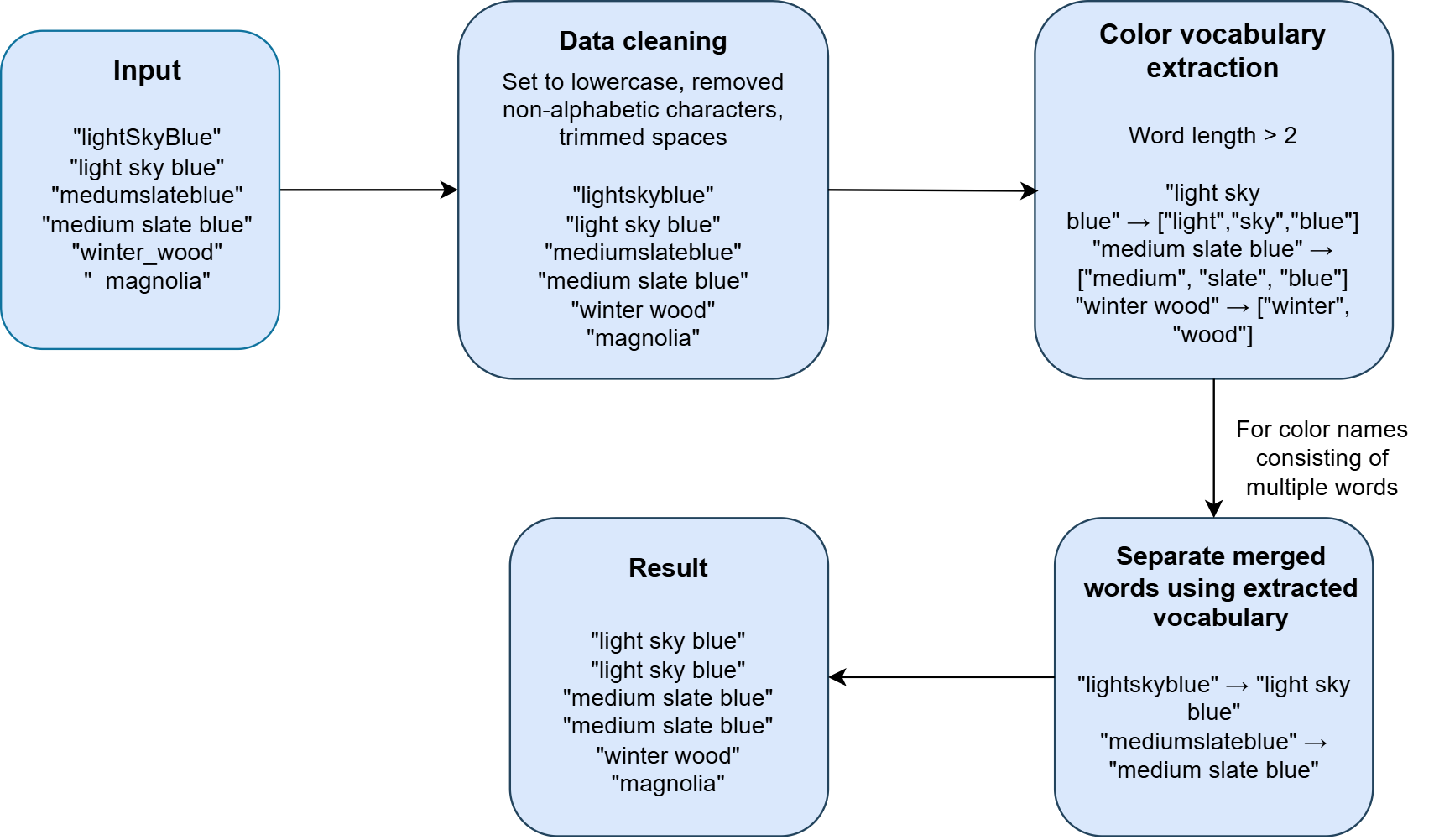}
\caption{Data preprocessing pipeline for color name normalization. The process includes converting to lowercase, removing non-alphabetic characters, trimming, and segmenting merged words. This ensures consistent formatting and correct identification of semantically equivalent color names (e.g., “lightskyblue” → “light sky blue”).}
    \label{fig:preprocess}
\end{figure*}

\begin{figure*}[h]
    \centering
    \includegraphics[width=0.8\textwidth]{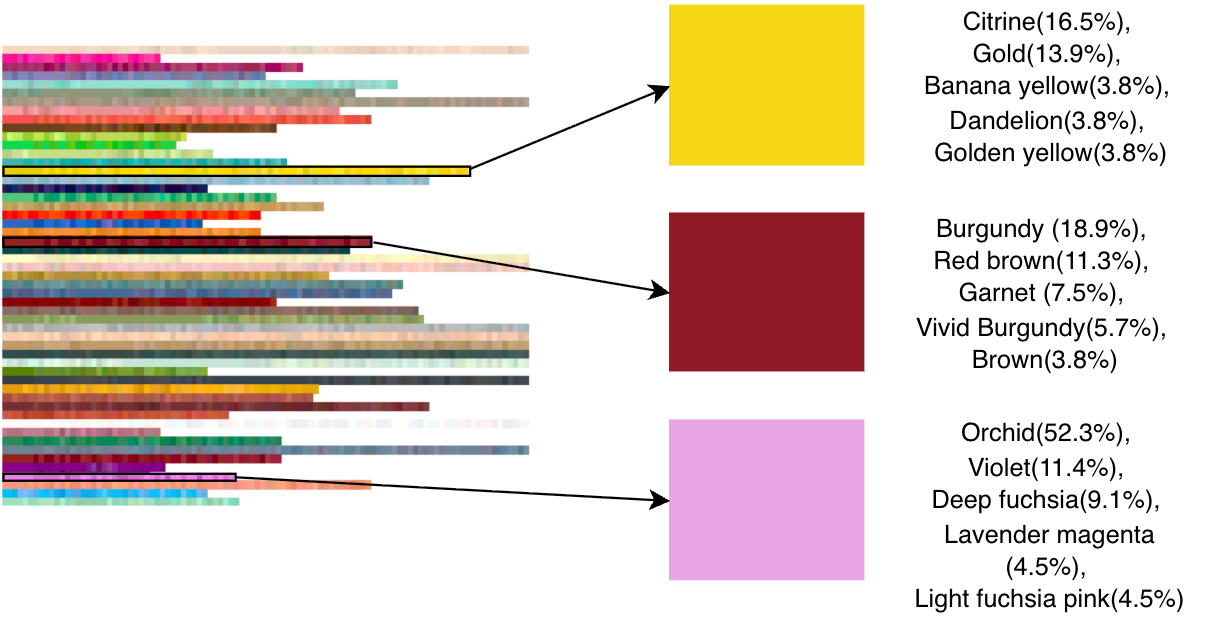}
     \caption{Examples of obtained clusters}
    \label{fig:clusters_with_examples}
\end{figure*}

\subsubsection{Data Collection}

\begin{table}[tb]
\centering
\caption{Number of Colors per Data Source}
\label{tab:color_sources}
\begin{tabular}{|p{5cm}|p{3cm}|}
\hline
\textbf{Source Name} & \textbf{Number of Colors} \\
\hline
Benjamin Moore & 3191  \\
Pantone & 2310 \\
Xona & 2031  \\
Dunnedwards Colors & 1847  \\
MatthewsPaint & 1727 \\
Clark Kensington & 1390 \\
ColorName & 1326 \\
Plochere & 1254 \\
ColorHexColors & 986 \\
ColorHexa & 746  \\
99 Colors & 672 \\
Figma & 432 \\
XKCD & 364  \\
HTML Colors & 363  \\
Crayola Colors & 169 \\
Magnolia Colors & 165  \\
W3Schools & 148 \\
Bokeh Colors & 148 \\
X11 & 145 \\
Kliz & 141 \\
\hline
Total & 19555\\
\hline
\end{tabular}
\end{table}
The dataset for this study was parsed from 20 diverse online sources, each contributing a different number of color entries as detailed in Table \ref{tab:color_sources}. This multi-source approach was chosen to capture the broad spectrum of color-naming conventions across domains, including design websites, color palette generators, paint manufacturer databases, and digital art platforms.

 Using automated web scraping, we collected a dataset of 19,555 color-name pairs. 
Each color entry in the final dataset contains five key attributes: the assigned color name, the hexadecimal color code, the corresponding RGB value, and the source. Representative examples of the collected data are presented in Table \ref{tab:example_of_color_data}.

\begin{table}[tb]
\centering
\caption{Examples of data in the collected dataset}
\label{tab:example_of_color_data}
\begin{tabular}{|l|l|l|l|}
\hline

\textbf{Name} & \textbf{Hex} & \textbf{RGB} & \textbf{Source} \\
\hline
snow white & \#f2f0eb & (242, 240, 235) & \url{https://margaret2.gi...} \\
\hline
vanilla ice & \#f0eada & (240, 234, 218) & \url{https://margaret2.gi...} \\
\hline
banana crepe & \#e7d3ad & (231, 211, 173) & \url{https://margaret2.gi...} \\
\hline
almond oil & \#f4efc1 & (244, 239, 193) & \url{https://margaret2.gi...} \\
\hline
angora & \#df1bb & (223, 209, 187) & \url{https://margaret2.gi...} \\
\hline
\end{tabular}
\end{table}


It is important to note that color naming is inherently influenced by language and culture, with existing models developed for Vietnamese, Italian, Persian and Russian, among others \cite{cr1,cr2,cr3, cr4}. However, this study focuses exclusively on English-language color naming.

\subsubsection{Data Preprocessing}

The raw dataset showed significant discrepancies in the formatting and structure of color names. To clean the data, excess spaces are trimmed, and regular expressions are used to remove any non-alphabetic characters. All color names were then transformed to lowercase. 

The preprocessing pipeline (see Fig. \ref{fig:preprocess}) transforms raw color names into standardized forms by applying a series of cleaning and segmentation steps, which are crucial for accurate name-frequency analysis.

A major challenge in the dataset was the presence of color names in both separated and merged forms, such as 'light sky blue' and 'lightskyblue.' To tackle this problem, we developed an extensive vocabulary of color-related terms by extracting individual words from all correctly formatted multi-word color names found in the dataset.

 The extraction method involved breaking the labels into constituent parts and applying a minimum character length filter of three to eliminate trivial words and common articles. The generated vocabulary served as a reference dictionary of all valid color expressions found throughout the dataset. For example, when the algorithm encountered 'lightskyblue', it could identify each component word ('light', 'sky', 'blue') by its inclusion in the vocabulary, thereby reconstructing the correctly spaced phrase 'light sky blue'. 



\subsection{K-means clustering}
K-Means clustering is an unsupervised learning technique widely used in color analysis. It groups visually similar pixel values. By partitioning the data into clusters based on similarity, the algorithm enables the extraction of representative dominant colors \cite{11139272, 10759889}. Given a set of color vectors

\begin{equation}
\mathbf{X} = \{x_1, x_2, \dots, x_n\}, \quad x_i \in \mathbb{R}^3
\end{equation}
where each vector represents the LAB components of a pixel, the objective of K-means is to partition the data into k clusters by minimizing the within-cluster sum of squared distances.
The clustering objective function is defined as:
\\
\begin{equation}
J = \sum_{i=1}^{k} \sum_{x \in C_i} \| x - \mu_i \|^2
\end{equation}
where $C_i$ denotes the set of points assigned to cluster $i$, and $\mu_i$ represents the centroid of that cluster.

\begin{algorithm}
\caption{Optimal Clustering using CIEDE2000 and Elbow Method}
\begin{algorithmic}[1]
\State \textbf{Input:} LAB color dataset $L$
\State \textbf{Output:} Optimal number of clusters $k^*$, cluster labels, RGB centroids

\State Define range of cluster numbers $K = \{50, 60, \ldots, 990\}$
\State Initialize empty list $\mathcal{S} \gets []$ \Comment{List of $\Delta E_{00}$ inertias}

\For{each $k \in K$}
    \State Run KMeans clustering with $k$ clusters on $L$
    \State Obtain labels and cluster centers $\mu_1, \ldots, \mu_k$
\State Initialize list $\mathcal{D} \gets []$ \Comment{Intra-cluster $\Delta E_{00}$s}

    \For{each cluster $C_i$}
        \State Extract points $P_i$ assigned to cluster $i$
        \If{$P_i$ is not empty}
            \State Compute CIEDE2000 distances between $P_i$ and $\mu_i$
            \State Append average distance to $\mathcal{D}$
        \EndIf
    \EndFor

    \State Append mean of $\mathcal{D}$ to $\mathcal{S}$
\EndFor

\State Plot curve $(k, \mathcal{S}_k)$ and identify elbow point $k^*$ using KneeLocator
\State Run KMeans with $k^*$ clusters to obtain final labels and centroids
\State Convert LAB centroids to RGB
\State Visualize clusters as color swatches using final labels
\end{algorithmic}
\end{algorithm}

K-means clustering was used to systematically group perceptually similar colors in the transformed CIELAB color space. CIELAB is a widely used perceptual color space in which equal numerical differences roughly correspond to equal perceived color differences \cite{muratbekova2025colormodelsimageprocessing}. CIELAB is mainly used for device‑independent color representation and quantization in imaging and graphics \cite{9945709, 11139184}.

Rather than arbitrarily selecting the number of clusters, an Elbow Method analysis was conducted across k values ranging from 50 to 1,000 to determine the optimal granularity.
The evaluation metric was the average intra-cluster perceptual difference, measured using CIEDE2000, which quantifies color differences as perceived by human vision rather than the simple Euclidean distance. The KneeLocator algorithm automatically identified the optimal number of clusters as the point at which increasing the cluster count yielded diminishing improvements in perceptual accuracy. Fig. \ref{fig:clusters_with_examples} illustrates the cluster examples. The number of clusters provided sufficient resolution to capture significant color differences while keeping the analysis of the following names manageable. The clustering process employed k-means++ initialization, which carefully selects the initial cluster centroids. 

As shown in Algorithm 1, it selects the optimal number of clusters for color data using the Elbow Method based on the perceptual color difference metric $\Delta E_{00}$ (CIEDE2000). For each $k$, it computes the average intra-cluster color difference and identifies the elbow point at which increasing $k$ yields minimal additional improvement. The optimal $k$ is then used for final K-Means clustering, and the resulting clusters are visualized using their RGB centroids.

\begin{figure}[tb]
    \centering
    \includegraphics[width=\linewidth]{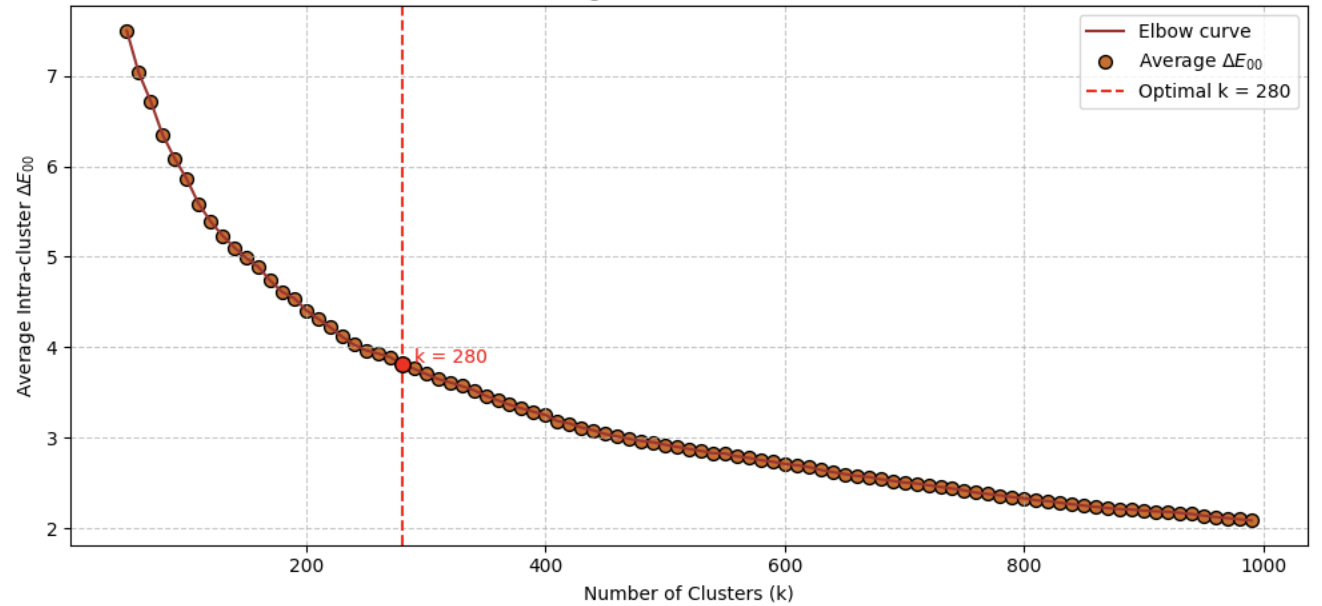}
    \caption{Elbow plot showing the average intra-cluster color difference ($\Delta E_{00}$) as a function of the number of clusters. The optimal number of clusters was determined to be $k = 280$.}
    \label{fig:elbow_plot}
\end{figure}

\begin{table*}[h]
\centering
\caption{Color names for some of the resulting RGB values}\label{tab:color_scheme}
\begin{tabular}{@{}ccc>{\centering\arraybackslash}p{7cm}@{}}
\toprule
\textbf{Color} & \textbf{RGB} & \textbf{Color names} \\
\midrule

\begin{tikzpicture}[baseline=2ex]
    \definecolor{fillcolor}{RGB}{135,196,111}
    \fill[fillcolor] (0,0) rectangle (0.7,0.7);
\end{tikzpicture}
& (135, 196, 111)
& pistachio, mantis, green palace, dollar bill, bud green pantone \\

\begin{tikzpicture}[baseline=2ex]
    \definecolor{fillcolor}{RGB}{245,6,244}
    \fill[fillcolor] (0,0) rectangle (0.7,0.7);
\end{tikzpicture}
& (245, 6, 244)
& magenta, neon pink, fuchsia, deep magenta, phlox \\

\begin{tikzpicture}[baseline=2ex]
    \definecolor{fillcolor}{RGB}{36,24,114}
    \fill[fillcolor] (0,0) rectangle (0.7,0.7);
\end{tikzpicture}
& (36, 24, 114)
& blue midnight, persian indigo, hippie blue, regimental, cosmic cobalt \\

\begin{tikzpicture}[baseline=2ex]
    \definecolor{fillcolor}{RGB}{205,75,139}
    \fill[fillcolor] (0,0) rectangle (0.7,0.7);
\end{tikzpicture}
& (205, 75, 139)
& mulberry, fuchsia purple, raspberry pink, magenta pantone, pink pantone \\

\begin{tikzpicture}[baseline=2ex]
    \definecolor{fillcolor}{RGB}{244,107,17}
    \fill[fillcolor] (0,0) rectangle (0.7,0.7);
\end{tikzpicture}
& (244, 107, 17)
& pumpkin, spanish orange, persimmon, safety orange, chocolate \\

\begin{tikzpicture}[baseline=2ex]
    \definecolor{fillcolor}{RGB}{196,128,65}
    \fill[fillcolor] (0,0) rectangle (0.7,0.7);
\end{tikzpicture}
& (196, 128, 65)
& copper, peru, tan, bakery brown, raw sienna \\

\begin{tikzpicture}[baseline=2ex]
    \definecolor{fillcolor}{RGB}{87,25,137}
    \fill[fillcolor] (0,0) rectangle (0.7,0.7);
\end{tikzpicture}
& (87, 25, 137)
& indigo, purple rebecca, dark orchid, purple, purple heart \\

\begin{tikzpicture}[baseline=2ex]
    \definecolor{fillcolor}{RGB}{143,25,38}
    \fill[fillcolor] (0,0) rectangle (0.7,0.7);
\end{tikzpicture}
& (143, 25, 38)
& burgundy, red brown, garnet, vivid burgundy, brown \\

\begin{tikzpicture}[baseline=2ex]
    \definecolor{fillcolor}{RGB}{205,33,57}
    \fill[fillcolor] (0,0) rectangle (0.7,0.7);
\end{tikzpicture}
& (205, 33, 57)
& crimson, rusty red, cherry, amaranth, cardinal \\

\begin{tikzpicture}[baseline=2ex]
    \definecolor{fillcolor}{RGB}{200,119,38}
    \fill[fillcolor] (0,0) rectangle (0.7,0.7);
\end{tikzpicture}
& (200, 119, 38)
& ochre, bronze, cinnamon, ginger, brown orange \\

\begin{tikzpicture}[baseline=2ex]
    \definecolor{fillcolor}{RGB}{60,4,252}
    \fill[fillcolor] (0,0) rectangle (0.7,0.7);
\end{tikzpicture}
& (60, 4, 252)
& blue, electric indigo, han purple, indigo, electric ultramarine \\

\begin{tikzpicture}[baseline=2ex]
    \definecolor{fillcolor}{RGB}{123,184,45}
    \fill[fillcolor] (0,0) rectangle (0.7,0.7);
\end{tikzpicture}
& (123, 184, 45)
& apple green, active green, jasmine green pantone, neon green cmyk, christi \\

\begin{tikzpicture}[baseline=2ex]
    \definecolor{fillcolor}{RGB}{75,90,34}
    \fill[fillcolor] (0,0) rectangle (0.7,0.7);
\end{tikzpicture}
& (75, 90, 34)
& dark green olive, army green, dark moss green, navy green, clove \\

\begin{tikzpicture}[baseline=2ex]
    \definecolor{fillcolor}{RGB}{232,166,227}
    \fill[fillcolor] (0,0) rectangle (0.7,0.7);
\end{tikzpicture}
& (232, 166, 227)
& plum, brilliant lavender, lavender rose, medium lavender magenta, orchid crayola \\

\bottomrule
\end{tabular}
\end{table*}

\subsection{Cluster–Name Association via Frequency Analysis}
We used a frequency-based aggregation technique based on actual naming data to assign representative color names to each cluster. First, each cluster centroid was converted to its corresponding RGB representation in the LAB color space. A precomputed mapping that links precise RGB colors to all human-provided color names observed for that color in the dataset used this RGB value as a key.

We obtained the whole collection of related color names for each cluster whose centroid RGB value was present in the mapping. To ensure that the naming statistics reflected direct human associations rather than assumed similarity across nearby regions of color space, these names were selected only from colors that precisely matched the centroid's RGB representation. 


We counted the frequency of all valid color names within each cluster and normalized the results by the total number of names associated with the cluster. This resulted in a percentage-based distribution showing the frequency with which each name was used to characterize that hue. The obtained values can be understood as empirical probabilities indicating how likely it is that a human observer would assign the cluster's representative hue a specific name. We kept only the top five most common names for each cluster.




\section{Results}
\subsection{Experimental Results}

\begin{figure}[tb]
    \centering
    \includegraphics[width=\linewidth]{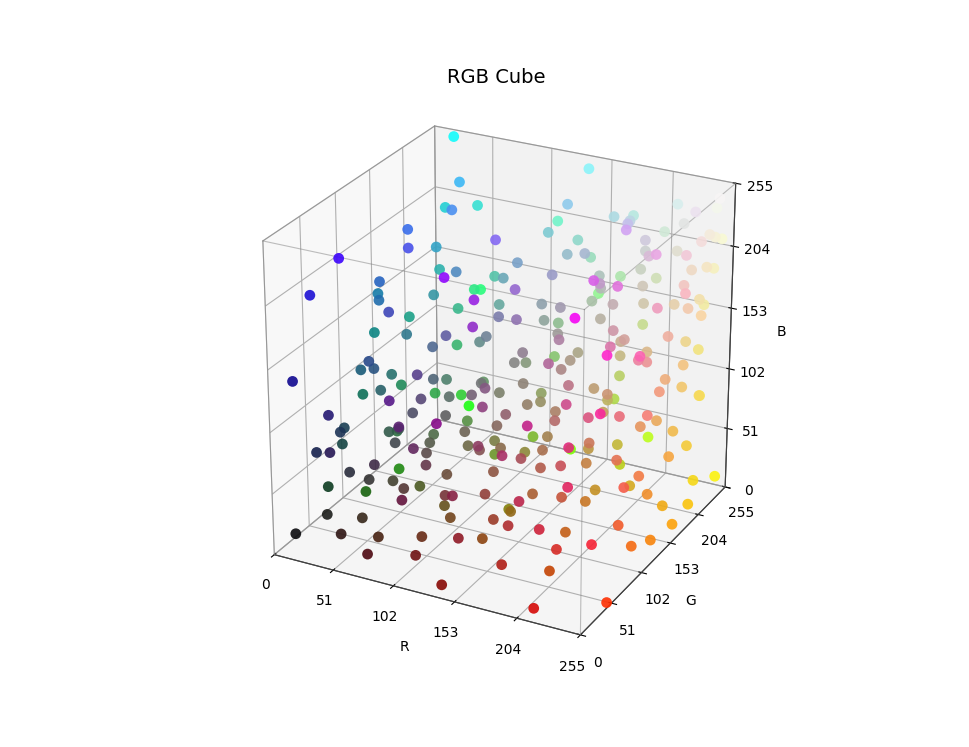}
    \caption{280 colors in RGB Cube}
    \label{fig:rgb_cube_colors}
\end{figure}

\begin{figure}[tb]
    \centering
    \includegraphics[width=\linewidth]{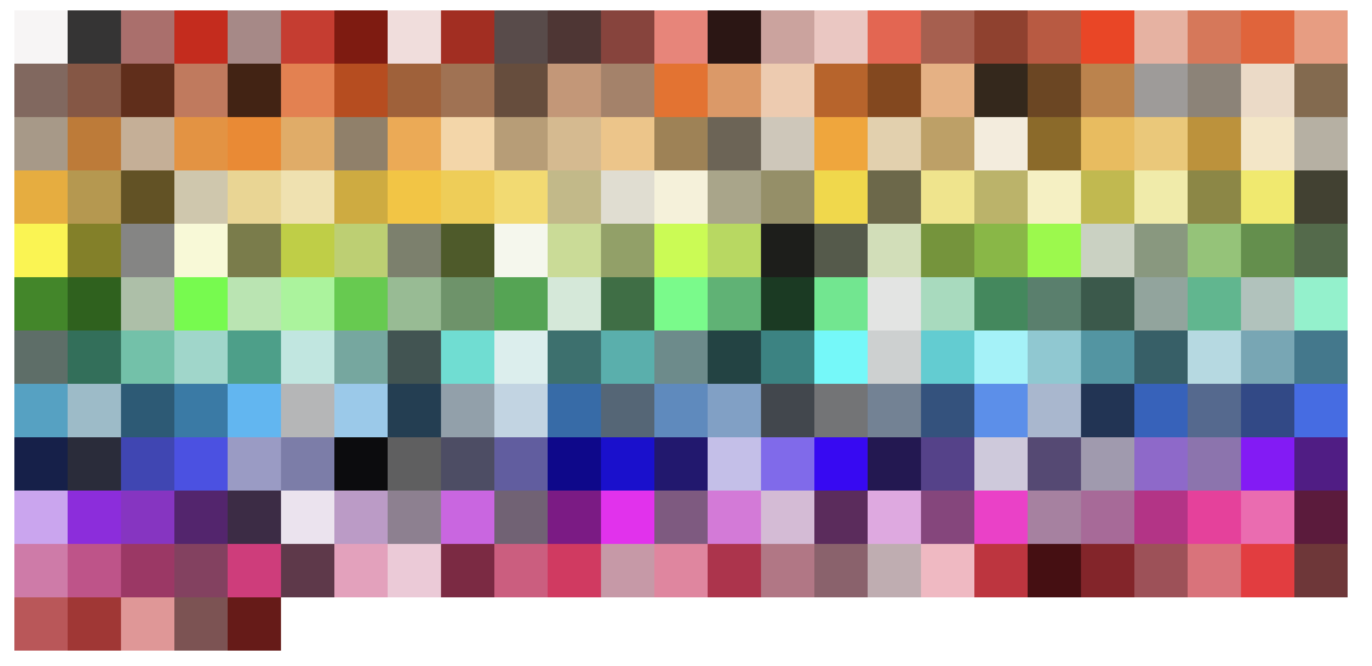}
    \caption{Grid visualization of obtained colors sorted by hue}
    \label{fig:grid_visualization}
\end{figure}

To determine the optimal number of clusters, we applied the Elbow Method using the perceptual color difference metric $\Delta E_{00}$. 
As shown in Fig.~\ref{fig:elbow_plot}, the curve begins to flatten around $k = 280$, which was identified as the optimal number of clusters using the KneeLocator algorithm.


\begin{figure*}[h]
    \centering
    \includegraphics[width=0.6\linewidth]{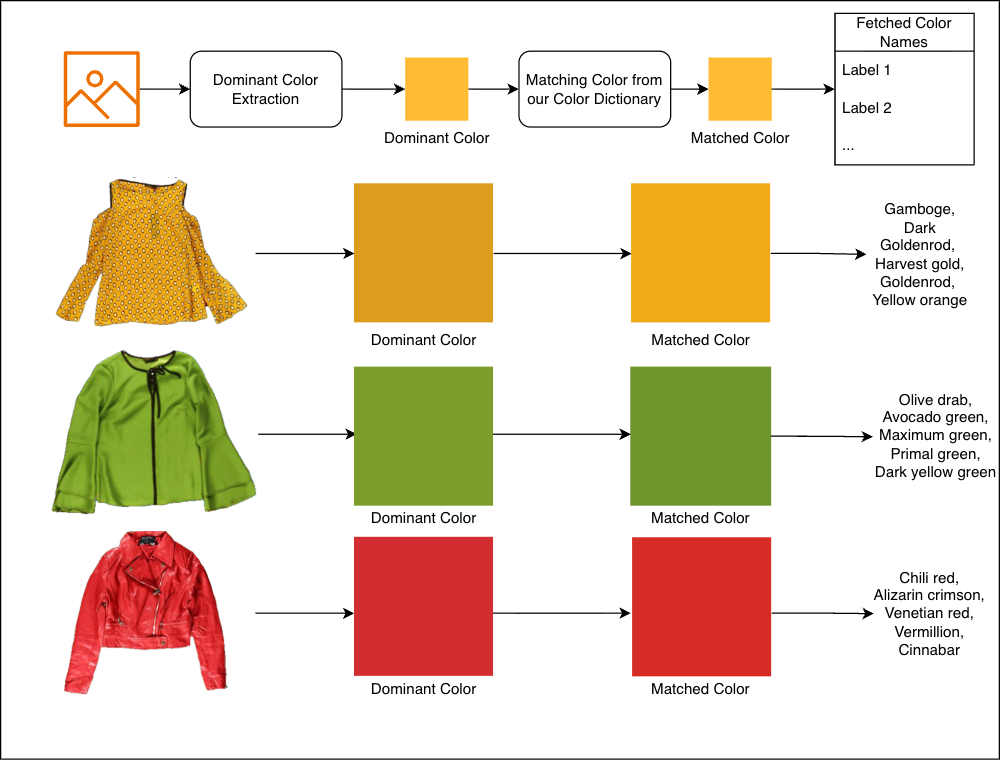}
     \caption{Image color tagging of Visuelle dataset}
    \label{fig:image_color_tagging}
\end{figure*}
The 280 resulting colors are displayed within the RGB color cube in Fig. \ref{fig:rgb_cube_colors} to visually represent their distribution. Furthermore, the variety of colors is depicted in Fig. \ref{fig:grid_visualization} in a grid format organized by hue.



 So, the K-means clustering algorithm effectively divided the 19,555 unique colors into 280 separate clusters in the LAB color space. Each cluster was characterized by its centroid coordinates, which were converted to RGB values.

As illustrated in Fig. \ref{fig:clusters_with_examples}, color names are attributed to the centroid of every cluster. These names are refined to the top five most probable selections, organized from highest to lowest probability.


The Table \ref{tab:color_scheme} presents representative examples of these cluster centroids, showing their corresponding RGB values along with the five most likely color names assigned to each centroid. It illustrates the connection between numerical color values and their semantic color names, emphasizing the variety of possible linguistic interpretations linked to each cluster. For instance, the RGB color (196, 128, 65) appears under multiple names across different sources - such as copper, peru, tan, bakery brown, and raw sienna. This illustrates the ambiguity present in current color naming practices, where a single color can carry numerous labels.

\subsection{Application Example in Content-based Image Retrieval}
The proposed color system has broad applicability for automatically labeling large image datasets and their descriptors.

The database could enhance image retrieval by content by using color as a retrieval criterion, enabling automatic image labeling and aiding in GAN-based text-to-image generation, consistent color tags.

To demonstrate the use of our color dataset, we applied it to the Visuelle\cite{Visuelle2021} dataset(see Fig. \ref{fig:image_color_tagging}) by first extracting each image’s dominant color using Color Thief and converting the resulting RGB values to Lab. We then searched our dataset for the closest matching color-cluster centroid and retrieved the top five associated color names for each dominant color.

VISUELLE\cite{Visuelle2021} is a publicly available multimodal dataset introduced for the task of forecasting sales of newly released fashion products. It is derived from real commercial data and comprises 5,577 new products sold across 100 retail stores between October 2016 and December 2019, accounting for approximately 45 million individual sales records. For each product, VISUELLE offers a rich set of annotations spanning visual, textual, temporal, and external popularity information, enabling comprehensive multimodal analysis. Each VISUELLE product is annotated with visual, textual, temporal, and external popularity information. High-quality RGB images are captured in a controlled studio environment with uniform white backgrounds and resolutions ranging from 256×256 to 1193×1172 pixels.

\section{Conclusion}

In this study, we proposed a multisource, data-driven framework for developing a standardized color-naming system by clustering 19,555 RGB–name pairs collected from 20 diverse sources. Using the CIELAB color space and the CIEDE2000 distance metric, we identified 280 color clusters and assigned representative names based on frequency analysis, focusing on linguistic patterns observed in real-world English usage. Our approach addresses key limitations in existing color standards, such as excessive granularity, inconsistent naming, and poor alignment with human perception.

This work contributes to the development of a practical solution for color labeling in design, digital media, content-based image retrieval, and generative AI. The system bridges the gap between machine-readable color codes and human-intuitive color descriptors. This research's implications include the potential cross-industry impact of establishing a standardized color-naming system.

While this study is limited to English, future work will explore multilingual extensions, incorporate human subject validation, and refine clustering methods using adaptive or learning-based techniques.

\bibliography{library}

@article{menegaz2006discrete,
  title={A discrete model for color naming},
  author={Menegaz, Gloria and Le Troter, Arnaud and Sequeira, Jean and Boi, Jean-Marc},
  journal={EURASIP Journal on Advances in Signal Processing},
  volume={2007},
  number={1},
  pages={029125},
  year={2006},
  publisher={Springer}
}

@book{lammens1994computational,
  title={A computational model of color perception and color naming},
  author={Lammens, Johan Maurice Gisele},
  year={1994},
  publisher={State University of New York at Buffalo}
}

@book{berlin1991basic,
  title={Basic color terms: Their universality and evolution},
  author={Berlin, Brent and Kay, Paul},
  year={1991},
  publisher={Univ of California Press}
}

@inproceedings{mylonas2010towards,
  title={Towards an online color naming model},
  author={Mylonas, Dimitris and MacDonald, Lindsay and Wuerger, Sophie},
  booktitle={Color and imaging conference},
  volume={18},
  pages={140--144},
  year={2010},
  organization={Society of Imaging Science and Technology}
}

@inproceedings{heer2012color,
  title={Color naming models for color selection, image editing and palette design},
  author={Heer, Jeffrey and Stone, Maureen},
  booktitle={Proceedings of the SIGCHI Conference on Human Factors in Computing Systems},
  pages={1007--1016},
  year={2012}
}

@article{kay1978linguistic,
  title={The linguistic significance of the meanings of basic color terms},
  author={Kay, Paul and McDaniel, Chad K},
  journal={Language},
  volume={54},
  number={3},
  pages={610--646},
  year={1978},
  publisher={Linguistic Society of America}
}

@inproceedings{moroney2008lexical,
  title={Lexical image processing},
  author={Moroney, Nathan and Obrador, Pere and Beretta, Giordano},
  booktitle={Color and Imaging Conference},
  volume={16},
  pages={268--273},
  year={2008},
  organization={Society of Imaging Science and Technology}
}

@article{benavente2008parametric,
  title={Parametric fuzzy sets for automatic color naming},
  author={Benavente, Robert and Vanrell, Maria and Baldrich, Ramon},
  journal={Journal of the Optical Society of America A},
  volume={25},
  number={10},
  pages={2582--2593},
  year={2008},
  publisher={Optical Society of America}
}

@article{seaborn2005fuzzy,
  title={Fuzzy colour category map for the measurement of colour similarity and dissimilarity},
  author={Seaborn, Matthew and Hepplewhite, Lee and Stonham, John},
  journal={Pattern Recognition},
  volume={38},
  number={2},
  pages={165--177},
  year={2005},
  publisher={Elsevier}
}

@inproceedings{yang1998visual,
  title={Visual tracking for multimodal human computer interaction},
  author={Yang, Jie and Stiefelhagen, Rainer and Meier, Uwe and Waibel, Alex},
  booktitle={Proceedings of the SIGCHI conference on Human factors in computing systems},
  pages={140--147},
  year={1998}
}

@book{plataniotis2000color,
  title={Color image processing and applications},
  author={Plataniotis, Konstantinos and Venetsanopoulos, Anastasios N},
  year={2000},
  publisher={Springer Science \& Business Media}
}

@article{gevers2000pictoseek,
  title={Pictoseek: Combining color and shape invariant features for image retrieval},
  author={Gevers, Theo and Smeulders, Arnold WM},
  journal={IEEE transactions on Image Processing},
  volume={9},
  number={1},
  pages={102--119},
  year={2000},
  publisher={IEEE}
}

@misc{muratbekova2025colormodelsimageprocessing,
      title={Color Models in Image Processing: A Review and Experimental Comparison}, 
      author={Muragul Muratbekova and Nuray Toganas and Ayan Igali and Maksat Shagyrov and Elnara Kadyrgali and Adilet Yerkin and Pakizar Shamoi},
      year={2025},
      eprint={2510.00584},
      archivePrefix={arXiv},
      primaryClass={cs.CV},
      url={https://arxiv.org/abs/2510.00584}, 
}

@INPROCEEDINGS{11139272,
  author={Sagatbek, Akmira and Seidakhmetova, Amina and Shamoi, Pakizar},
  booktitle={2025 IEEE 5th International Conference on Smart Information Systems and Technologies (SIST)}, 
  title={Comparative Analysis of Clustering Algorithms for Human-Consistent Dominant Color Extraction}, 
  year={2025},
  volume={},
  number={},
  pages={1-7},
  doi={10.1109/SIST61657.2025.11139272}}

@INPROCEEDINGS{10759889,
  author={Shagyrov, Maksat and Shamoi, Pakizar},
  booktitle={2024 Joint 13th International Conference on Soft Computing and Intelligent Systems and 25th International Symposium on Advanced Intelligent Systems (SCIS\&ISIS)}, 
  title={Color and Sentiment: A Study of Emotion-Based Color Palettes in Marketing}, 
  year={2024},
  volume={},
  number={},
  pages={1-7},
  doi={10.1109/SCISISIS61014.2024.10759889}}

@INPROCEEDINGS{11139184,
  author={Burambekova, Aruzhan and Shamoi, Pakizar},
  booktitle={2025 IEEE 5th International Conference on Smart Information Systems and Technologies (SIST)}, 
  title={Comparative Analysis of Color Models for Human Perception and Visual Color Difference}, 
  year={2025},
  volume={},
  number={},
  pages={1-6},
  doi={10.1109/SIST61657.2025.11139184}}

@INPROCEEDINGS{9945709,
  author={Shamoi, Pakizat and Sansyzbayev, Daniyar and Abiley, Nurmukhamed},
  booktitle={2022 International Conference on Smart Information Systems and Technologies (SIST)}, 
  title={Comparative Overview of Color Models for Content-Based Image Retrieval}, 
  year={2022},
  volume={},
  number={},
  pages={1-6},
  doi={10.1109/SIST54437.2022.9945709}}

@INPROCEEDINGS{10001872,
  author={Shamoi, Pakizar and Inoue, Atsushi and Kawanaka, Hiroharu},
  booktitle={2022 Joint 12th International Conference on Soft Computing and Intelligent Systems and 23rd International Symposium on Advanced Intelligent Systems (SCIS\&ISIS)}, 
  title={Color Aesthetics and Context-Dependency}, 
  year={2022},
  volume={},
  number={},
  pages={1-7},
  doi={10.1109/SCISISIS55246.2022.10001872}}

@article{gevers1999color,
  title={Color-based object recognition},
  author={Gevers, Theo and Smeulders, Arnold WM},
  journal={Pattern recognition},
  volume={32},
  number={3},
  pages={453--464},
  year={1999},
  publisher={Elsevier}
}

@article{li2019controllable,
  title={Controllable text-to-image generation},
  author={Li, Bowen and Qi, Xiaojuan and Lukasiewicz, Thomas and Torr, Philip},
  journal={Advances in neural information processing systems},
  volume={32},
  year={2019}
}

@article{feuerriegel2024generative,
  title={Generative ai},
  author={Feuerriegel, Stefan and Hartmann, Jochen and Janiesch, Christian and Zschech, Patrick},
  journal={Business \& Information Systems Engineering},
  volume={66},
  number={1},
  pages={111--126},
  year={2024},
  publisher={Springer}
}

@article{cr4,
author = {Jameson, Kimberly A. and Alvarado, Nancy},
title = {Differences in color naming and color salience in Vietnamese and English},
journal = {Color Research \& Application},
volume = {28},
number = {2},
pages = {113-138},
doi = {https://doi.org/10.1002/col.10131},
url = {https://onlinelibrary.wiley.com/doi/abs/10.1002/col.10131},
eprint = {https://onlinelibrary.wiley.com/doi/pdf/10.1002/col.10131},
year = {2003}
}

@article{cr5,
  title={What should we call this color? The influence of color-naming on consumers' attitude toward the product},
  author={Chou, Hsuan-Yi and Chu, Xing-Yu and Chiang, Yu-Han},
  journal={Psychology \& Marketing},
  volume={37},
  number={7},
  pages={942--960},
  year={2020},
  publisher={Wiley Online Library}
}

@article{cr3,
author = {Paramei, Galina V. and Griber, Yulia A. and Mylonas, Dimitris},
title = {An online color naming experiment in Russian using Munsell color samples},
journal = {Color Research \& Application},
volume = {43},
number = {3},
pages = {358-374},
doi = {https://doi.org/10.1002/col.22190},
url = {https://onlinelibrary.wiley.com/doi/abs/10.1002/col.22190},
eprint = {https://onlinelibrary.wiley.com/doi/pdf/10.1002/col.22190},
year = {2018}
}

@article{cr2,
author = {Kandi, S. Gorji and Tehran, M. Amani and Hassani, N. and Jarrahi, A.},
title = {Color naming for the Persian language},
journal = {Color Research \& Application},
volume = {40},
number = {4},
pages = {352-360},
doi = {https://doi.org/10.1002/col.21887},
url = {https://onlinelibrary.wiley.com/doi/abs/10.1002/col.21887},
eprint = {https://onlinelibrary.wiley.com/doi/pdf/10.1002/col.21887},
year = {2015}
}

@article{cr1,
author = {Paggetti, Giulia and Menegaz, Gloria and Paramei, Galina V.},
title = {Color naming in Italian language},
journal = {Color Research \& Application},
volume = {41},
number = {4},
pages = {402-415},
doi = {https://doi.org/10.1002/col.21953},
url = {https://onlinelibrary.wiley.com/doi/abs/10.1002/col.21953},
eprint = {https://onlinelibrary.wiley.com/doi/pdf/10.1002/col.21953},
year = {2016}
}

@inproceedings{smith1997visualseek,
  title={VisualSEEk: a fully automated content-based image query system},
  author={Smith, John R and Chang, Shih-Fu},
  booktitle={Proceedings of the fourth ACM international conference on Multimedia},
  pages={87--98},
  year={1997}
}

@article{elliot2014color,
  title={Color psychology: Effects of perceiving color on psychological functioning in humans},
  author={Elliot, Andrew J and Maier, Markus A},
  journal={Annual review of psychology},
  volume={65},
  number={1},
  pages={95--120},
  year={2014},
  publisher={Annual Reviews}
}

@book{nassau2001physics,
  title={The physics and chemistry of color: the fifteen causes of color},
  author={Nassau, Kurt},
  year={2001}
}

@article{boynton1987locating,
  title={Locating basic colors in the OSA space},
  author={Boynton, Robert M and Olson, Conrad X},
  journal={Color Research \& Application},
  volume={12},
  number={2},
  pages={94--105},
  year={1987},
  publisher={Wiley Online Library}
}

@article{berk1982human,
  title={A human factors study of color notation systems for computer graphics},
  author={Berk, Toby and Kaufman, Arie and Brownston, Lee},
  journal={Communications of the ACM},
  volume={25},
  number={8},
  pages={547--550},
  year={1982},
  publisher={ACM New York, NY, USA}
}

@article{mojsilovic2005computational,
  title={A computational model for color naming and describing color composition of images},
  author={Mojsilovic, Aleksandra},
  journal={IEEE Transactions on Image processing},
  volume={14},
  number={5},
  pages={690--699},
  year={2005},
  publisher={IEEE}
}

@article{ozturk2005,
title={Location of Munsell colors in the RAL Design System},
author={L. Öztürk},
journal={Color Research and Application},
year={2005},
volume={30},
pages={130-134},
doi={10.1002/col.20091}
}

@article{VanDeWeijer2009Learning,
title={Learning Color Names for Real-World Applications},
author={Joost van de Weijer and C. Schmid and J. Verbeek and Diane Larlus},
journal={IEEE Transactions on Image Processing},
year={2009},
volume={18},
pages={1512-1523},
doi={10.1109/tip.2009.2019809}
}

@article{Hard1981NCS—Natural,
title={NCS—Natural Color System: A Swedish Standard for Color Notation},
author={A. Hård and Lars Sivik},
journal={Color Research and Application},
year={1981},
volume={6},
pages={129-138},
doi={10.1002/col.5080060303}
}

@article{Kelly2018Color,
title={Color: Universal Language and Dictionary of Names},
author={K. L. Kelly and D. B. Judd},
year={2018},
doi={}
}

@unknown{Visuelle2021,
author = {Skenderi, Geri and Joppi, Christian and Denitto, Matteo and Cristani, Marco},
year = {2021},
month = {09},
pages = {},
title = {Well Googled is Half Done: Multimodal Forecasting of New Fashion Product Sales with Image-based Google Trends},
doi = {10.48550/arXiv.2109.09824}
}

@article{Baronchelli2009Modeling,
title={Modeling the emergence of universality in color naming patterns},
author={Andrea Baronchelli and T. Gong and A. Puglisi and V. Loreto},
journal={Proceedings of the National Academy of Sciences},
year={2009},
volume={107},
pages={2403 - 2407},
doi={10.1073/pnas.0908533107}
}

@article{Lindsey2006Universality,
title={Universality of color names},
author={D. Lindsey and Angela M. Brown},
journal={Proceedings of the National Academy of Sciences},
year={2006},
volume={103},
pages={16608 - 16613},
doi={10.1073/pnas.0607708103}
}

@article{Loreto2012On,
title={On the origin of the hierarchy of color names},
author={V. Loreto and Animesh Mukherjee and F. Tria},
journal={Proceedings of the National Academy of Sciences},
year={2012},
volume={109},
pages={6819 - 6824},
doi={10.1073/pnas.1113347109}
}

@article{Kay2003Resolving,
title={Resolving the question of color naming universals},
author={P. Kay and T. Regier},
journal={Proceedings of the National Academy of Sciences of the United States of America},
year={2003},
volume={100},
pages={9085 - 9089},
doi={10.1073/pnas.1532837100}
}

@book{judd1975color,
  title={Color in Business, Science, and Industry},
  author={Judd, Deane B. and Wyszecki, G{\"u}nter},
  edition={3rd},
  year={1975},
  publisher={John Wiley \& Sons},
  address={New York},
  isbn={9780471452126}
}

@article{hardin1992virtues,
  title={The virtues of illusion},
  author={Hardin, C. L.},
  journal={Philosophical Studies},
  volume={68},
  number={3},
  pages={371--382},
  year={1992},
  publisher={Springer},
  doi={10.1007/BF00694852}
}

@book{goldstein1996sensation,
  title={Sensation and Perception},
  author={Goldstein, E. Bruce},
  edition={4th},
  year={1996},
  publisher={Brooks/Cole Publishing Co.},
  address={Pacific Grove, CA}
}

\end{document}